\documentclass[aps,twocolumn,showpacs,superscriptaddress]{revtex4-1}
\usepackage{amssymb}
\usepackage{amsmath}
\usepackage{graphicx}
\usepackage{hyperref} 
\usepackage[normalem]{ulem}  
\hypersetup{colorlinks=true,
            linkcolor=blue,
            citecolor=blue,
            urlcolor=orange}
        
\usepackage[dvipsnames]{xcolor}

\newcommand{\kv}{\ensuremath{\mathbf{k}}}

\begin{document}

\title{
The effect of local magnetic moments on spectral properties and resistivity \texorpdfstring{\\near}{Lg} the interaction- and doping induced Mott transitions}
\author{T. B. Mazitov}
\affiliation{Center for Photonics and 2D Materials, Moscow Institute of Physics and Technology, Institutsky lane 9, Dolgoprudny,
141700, Moscow region, Russia}
\author{A. A. Katanin}
\affiliation{Center for Photonics and 2D Materials, Moscow Institute of Physics and Technology, Institutsky lane 9, Dolgoprudny,
141700, Moscow region, Russia}
\affiliation{M. N. Mikheev Institute of Metal Physics, Kovalevskaya Street 18, 620219
Ekaterinburg, Russia.}

\begin{abstract}
We study the effect of the formation and screening of local magnetic moments on the temperature- and interaction dependencies of spectral functions and resistivity in the vicinity of the metal-insulator transition. 
We use  dynamical mean-field theory for the strongly correlated Hubbard model and associate the peculiarities of the above mentioned properties with those found for the local charge $\chi_c$ and spin $\chi_s$ susceptibilities. We show that at half filling the maximum of resistivity at a certain temperature $T^*$ corresponds to the appearance of central quasiparticle peak of the spectral function and  entering the metallic regime with well defined fermionic quasiparticles. At the same time, the temperature of the crossover to the regime with screening of local magnetic moments, determined by the minimum of double occupation, is smaller than the temperature scale $T^*$ and coincides at half filling with the boundary $T_{\beta=1}(U)$ corresponding to the exponent of resistivity $\beta\equiv d\ln \rho/d\ln T=1$.
Away from half filling we find weak increase of the temperature of the beginning and completion of the screening (i.e. Kondo temperature) of local magnetic moments, while the unscreened local magnetic moments exist only up to few percents of doping. 
In the low temperature regime $T<T_{\beta=1}$ simultaneous presence of itinerant and localized degrees of freedom yields almost linear temperature dependence of scattering rate and resistivity.
\end{abstract}

\maketitle


\section{Introduction}

The Mott metal-insulator transition (MIT) \cite{Mott} represents a  phenomenon, which occurs due to strong electronic correlations. It is observed in particular in transition metal oxides, such as V$_2$O$_3$ \cite{V2O3-1,V2O3-2,V2O3-3,V2O3-4,V2O3-5,V2O3-6,V2O3-7}, layered organic compounds \cite{Org1,Org2,Org3,Org4,Org5,OrgQC,OrgQP}, high-temperature cuprate superconductors (see, e.g. Ref. \cite{HighTc}), etc. The proximity to Mott transition yields peculiarities of many physical properties, which can be related to appearance of local magnetic moments (see, e.g., the discussion in Refs. \cite{Spalek,Nozieres,Fabrizio}). 

Quantitative studies of the Mott transition became possible with the discovery of the dynamical mean-field theory (DMFT) \cite{DMFT}. Originally, Mott transition was 
 described mainly on the basis of single-particle properties, e.g., spectral functions, densities of states, etc \cite{DMFT}.
The peculiarities of transport properties near Mott transition were investigated in Refs. \cite{Mott_res1,Mott_res1a,Mott_res2,Mott_res3,QC1, QC2, QC3, QC4,Widom1}. In this respect, the boundary of the metallic behavior (the so called Brinkmann-Rice boundary), which can be determined from the maximum of the temperature dependence of resistivity, was introduced \cite{OrgQP,Org3,Org4}. At the same time, the crossover between different states near the first order transition can be characterized by the Widom line, which was first introduced in the context of liquid-gas transition \cite{Widom,WidomLG} as the line of a sharp change from the liquid to gas-like behavior above the critical temperature. 
For correlated electronic systems this boundary, corresponding to a sharp change from metal to insulating behavior, can be determined in particular from the inflection points of the $\rho(U)$ dependencies \cite{OrgQC,QC2}, double occupation or single-particle properties \cite{Widom1}. The relation of the Widom line to the second derivative of the Landau functional, determined by the convergence rate of DMFT solutions, was emphasized in Refs. \cite{QC1,QC2}. Moreover, it was shown that near Mott transition the interaction dependence of the resistivity \cite{QC1,QC2,QC3,QC4}, the spectral functions and the self-energies \cite{QC4} obey certain scaling laws, showing quantum critical behavior of these quantities. 

Recently, in spite of studying the behavior of the two-particle quantities near Mott transition, the formation of local magnetic moments (LMM) in the vicinity of Mott transition  \cite{Toschi,Katsnelson,OurMott} and explicit determination of Landau functional and its derivatives in terms of the local vertices \cite{OurMott1} was discussed at half filling. In particular, criteria for determining the temperatures of the formation \cite{Katsnelson,OurMott}, start of the screening \cite{OurMott}, and fully screened (Kondo temperature) \cite{Toschi,Katsnelson,OurMott} local magnetic moments based on the peculiarities of the local charge and spin susceptibilities were proposed. The relation of these criteria to the above mentioned earlier discussed features of quantum critical behavior of resistivity near Mott transition is however not obvious, since it involves the relation {of} single-particle, {local} two-particle {and} transport properties. Although at the level of DMFT for the single-band model the vertex corrections to transport properties vanish and the two-particle properties are not directly related to transport properties, they are both related to the single-particle properties.

The resistivity at finite doping was studied previously within DMFT and cluster theories in Refs. \cite{Mott_res1a,QC3,Triangular1}. While close to half filling the temperature dependence of the resistivity shows a maximum, which is similar to that for the half filled case, the temperature interval of pronounced almost linear temperature dependence of resistivity is obtained in this case. Again, the question of the connection of this behavior to peculiarities of single-particle quantities can be raised in this case.

In the present paper we study a connection of the above described features of resistivity to the {local} single- and two particle properties. In particular, we supplement previous study of the formation of local magnetic moments at half filling \cite{OurMott} by detail study of the temperature- and interaction dependence of resistivity within the Hubbard model on the square lattice. By comparing the obtained characteristic temperatures at half filling we find that different criteria for the Widom line $T^*$ almost coincide with each other and with the start of the formation of quasiparticles. We also find the coincidence of the line of the resistivity exponent $\beta\equiv d\ln\rho/d\ln T=1$ with the boundary of start of the screening of local magnetic moments, determined previously from the minimum of local charge compressibility and double occupation. This provides natural definition of this boundary in terms of directly measurable quantities and allows us to conclude that the temperature of screening of LMM is located below $T^*$. Away from half filling we show presence of screened LMM. The boundary of the screening regime is determined by the minimum of double occupation, while the minimum of local charge compressibility disappears at sufficiently large doping due to presence of free charge carriers. We also show that the exponent of the resistivity $\beta=1$ 
corresponds to the onset of the linear temperature dependence of resistivity and quasiparticle damping in this case.

The plan of the paper is the following. In Sect. II we discuss briefly the model and method, in Sect III we present the results for half filling ($n=1$, Sect. IIIA) and away from half filling ($n<1$, Sect. IIIB). In Sect. IV we present conclusions.

\section{The model and method}

We consider the Hubbard model on the square lattice 
\begin{equation} \label{eq:hubbard}
H = - t\sum_{\langle i, j \rangle , \sigma}{c_{i\sigma}^\dagger c_{j\sigma}} + U\sum_i{n_{i\uparrow}n_{i\downarrow}},
\end{equation}
and use the half bandwidth $D=4t=1$ as the unit of energy.

We apply the DMFT approach \cite{DMFT} and evaluate the self-energies $\Sigma(\nu)$ and respective local spectral functions $A(\nu)$ at the real frequency axis using numerical renormalization group (NRG) approach \cite{NRG} within the TRIQS-NRG Ljubljana interface package \cite{TRIQS}.  Using DMFT results we evaluate the conductivity starting from Kubo formula by performing analytical continuation of the current-current correlation function
\begin{equation}
\sigma(i\omega_n) = \frac{2e^2 T}{\omega_n }\sum_{\textbf{k},\nu_n}v_{{\bf k},a}^2 G(\epsilon_\textbf{k}, i\nu_n)G(\epsilon_\textbf{k}, i\nu_n + i\omega_n)
\end{equation}
where $G(\epsilon,i\nu_n)=1/(i\nu_n-\epsilon-\Sigma(i\nu_n))$ are the Green's functions, $\epsilon_{\bf k}=-(\cos k_x+\cos k_y)/2$ is the dispersion, $v_{\bf k}=\nabla \epsilon_{\bf k}=(\sin k_x,\sin k_y)/2$ is the electron velocity, $\nu_n$ and $\omega_n$ are the fermionic and bosonic Matsubara frequencies, the factor of $2$ comes from the spin summation, and we neglect the vertex corrections, which vanish in the DMFT for the single-band model. 
Introducing partial density of states
\begin{align}
    {\mathcal D}_0(\epsilon)&=2\sum_{\textbf{k}} \sin^2 k_x \delta(\epsilon-\epsilon_{\bf k})\notag\\
    &=\frac{8}{\pi^2} \left[E(1-\epsilon^2)-\epsilon^2 K(1-\epsilon^2)\right],
\end{align}
where $E(x)$ and $K(x)$ 
are the complete elliptic integrals of the first and second kind, we find 
\begin{equation}
\sigma(i\omega_n) = \frac{e^2 T}{4\omega_n}\sum_{\nu_n}\int^{1}_{-1}d\epsilon {\mathcal D}_0(\epsilon)G(\epsilon, i\nu_n)G(\epsilon, i\nu_n + i\omega_n).
\end{equation}
Using spectral representation, 
performing summation over Matsubara frequencies, analytic continuation to the real axis $i\omega_n \rightarrow \omega + i\delta$, and taking the real part of the
the limit $\sigma(\omega \rightarrow 0)$ we finally obtain the static conductivity
\begin{eqnarray}
\sigma = -\frac{\pi e^2}{4}\int^{1}_{-1}d\epsilon {\mathcal D}_0(\epsilon)\int^{+\infty}_{-\infty}d\nu f'(\nu) A(\epsilon, \nu)^2, 
\label{eq:sigma}
\end{eqnarray}
where $A(\epsilon,\nu)=-(1/\pi){\rm Im}G(\epsilon,\nu)$ 
is the spectral density.
The position of the Widom line of the crossover from incoherent metal to insulating regime is determined from the maxima of $\rho(T)$ dependence \cite{OrgQP,Org3,Org4} or inflection points of $\ln \rho(U)$ dependence \cite{OrgQC,QC2}; we show below that these two criteria agree well with each other. 

{We are interested in the temperature range $T>T_K$, where $T_K$ is the Kondo temperature of screening of local magnetic moments \cite{OurMott,Pruschke,Mott_res1a}. Therefore, the van Hove singularity of the density of states is cut by temperature and quasiparticle damping, and we expect the obtained results to be not specific for the two-dimensional square lattice, but qualitatively applicable for the other forms of the bare density of states not gaped at the Fermi level (see, e.g., the discussion in Ref. \cite{Log}).}

At half filling, we also compare the position of the obtained Widom line to that found from the minimum of the lowest eigenvalue of the second derivative of the Landau functional $\Omega$ \cite{OurMott1}
\begin{align}
    \frac{\delta^2 \Omega}{\delta (i\Delta_\nu) \delta (i \Delta_{\nu'})}=
        \frac{\hat{1}}{\hat{1}-\hat{x}^{-1}\hat{X}}
    \left(\hat{1}-\hat{\mathcal{D}}\right)\hat{x}.
    \label{eq:OmegaStable}
\end{align}
where
$\Delta_\nu$ is the hybridization function of the impurity problem, $\mathcal{D}_{\nu \nu ^{\prime }}=\left(x_\nu-X_\nu\right) F^{\rm loc}_{\omega=0,\nu \nu ^{\prime }}$, 
$\hat{X}_{\nu}=- \hat{1}T\sum_\kv G(\epsilon_{\bf k},\nu)^2$ and $\hat{x}_{\nu}=-\hat{1}T G_{\rm loc}^2(\nu)$ are the non-local and local bubbles, considered as diagonal in frequency operators ($\hat{1}=\delta_{\nu\nu'}$), $G_{\rm loc}(\nu)$ is the local Green's function, $F_{\omega=0,\nu\nu'}^{\rm loc}$ is the zero bosonic frequency local vertex.
The local vertices are evaluated by using the {continuous-time} quantum Monte Carlo (CT-QMC) impurity solver, implemented in the 
iQIST software package \cite{iQIST}.

\section{Results}

\subsection{Half filling \texorpdfstring{$n=1$}{Lg}}
\label{Sectn1}

\begin{figure}[b]
\includegraphics[width=1.0\linewidth]{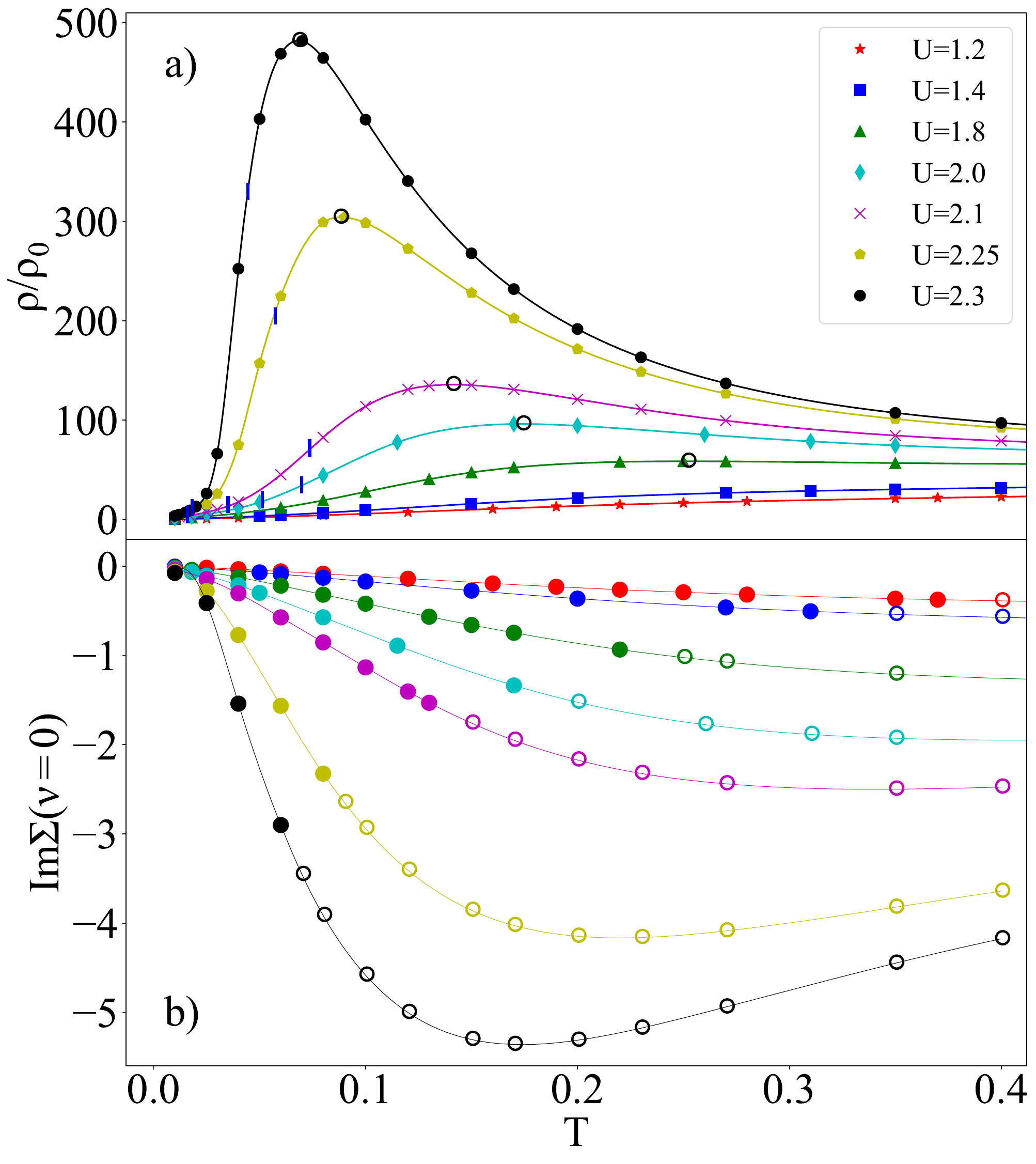}
\caption{(Color online). Temperature dependencies of (a) the resistivity (in units of $\rho_0=\hbar/(4e^2)$), (b) the imaginary part of zero-frequency self-energy at various values of $U$. The black open circles on (a) indicate maxima of $\rho$, the blue dashes on (a) indicate points of $\beta=d\ln\rho/d\ln T=1$. Filled (open) circles on (b) indicate negative (positive) sign of $\partial {\rm Re} \Sigma(\nu)/\partial\nu$.}
\label{fig:resistivity}
\end{figure}


We consider first the peculiarities of the resistivity near Mott transition at half filling. The temperature dependence of resistivity, calculated according to the Eq. (\ref{eq:sigma}) for various values of Coulomb interaction is shown in the Fig. \ref{fig:resistivity}a. In agreement with the previous studies on the infinite dimensional hypercubic \cite{Mott_res1,Mott_res1a,Mott_res2} and the square \cite{Mott_res3} lattice, the obtained dependencies show a maximum at some characteristic temperature $T^*(U)$, which decreases with an increase of the interaction strength. It  was suggested (see also experimental studies of layered organic compounds \cite{OrgQP,Org3,Org4}) that the maximum of the dependence $\rho(T)$ is related to the loss of quasiparticles with increasing temperature.

To confirm the relation of the obtained maxima of the resistivity to the loss of fermionic quasiparticles, we consider the temperature evolution of the electronic self-energy. Fig. \ref{fig:resistivity}b shows the temperature dependencies of ${\rm Im}\Sigma(\nu=0)$, at various interaction strength. Closed (open) circles correspond to the negative (positive) derivative $\partial {\rm Re}\Sigma(\nu)/\partial\nu$ and minimum (maximum) of $|{\rm Im}\Sigma(\nu)|$ at $\nu=0$, 
corresponding to well-defined (destroyed) fermionic quasiparticles. As one can see, the change of the sign of the derivative occurs indeed at approximately the same temperatures, as obtained from the maxima of the resistivity. 
We have also compared the obtained temperature $T^*$ to that of the change of the frequency dependence of the local spectral function $A(\nu)=-{\rm Im}G_{\rm loc}(\nu)/\pi$ \cite{OurMott}.
We have found that the appearance of the central quasiparticle peak of $A(\nu)$ also occurs at approximately the same temperature $T^*$, as the maximum of resistivity and onset of quasiparticle shape of the electronic self-energy (this correspondence holds also for the other values of $U$, see below).


\begin{figure}[t]
\includegraphics[width=1.0\linewidth]{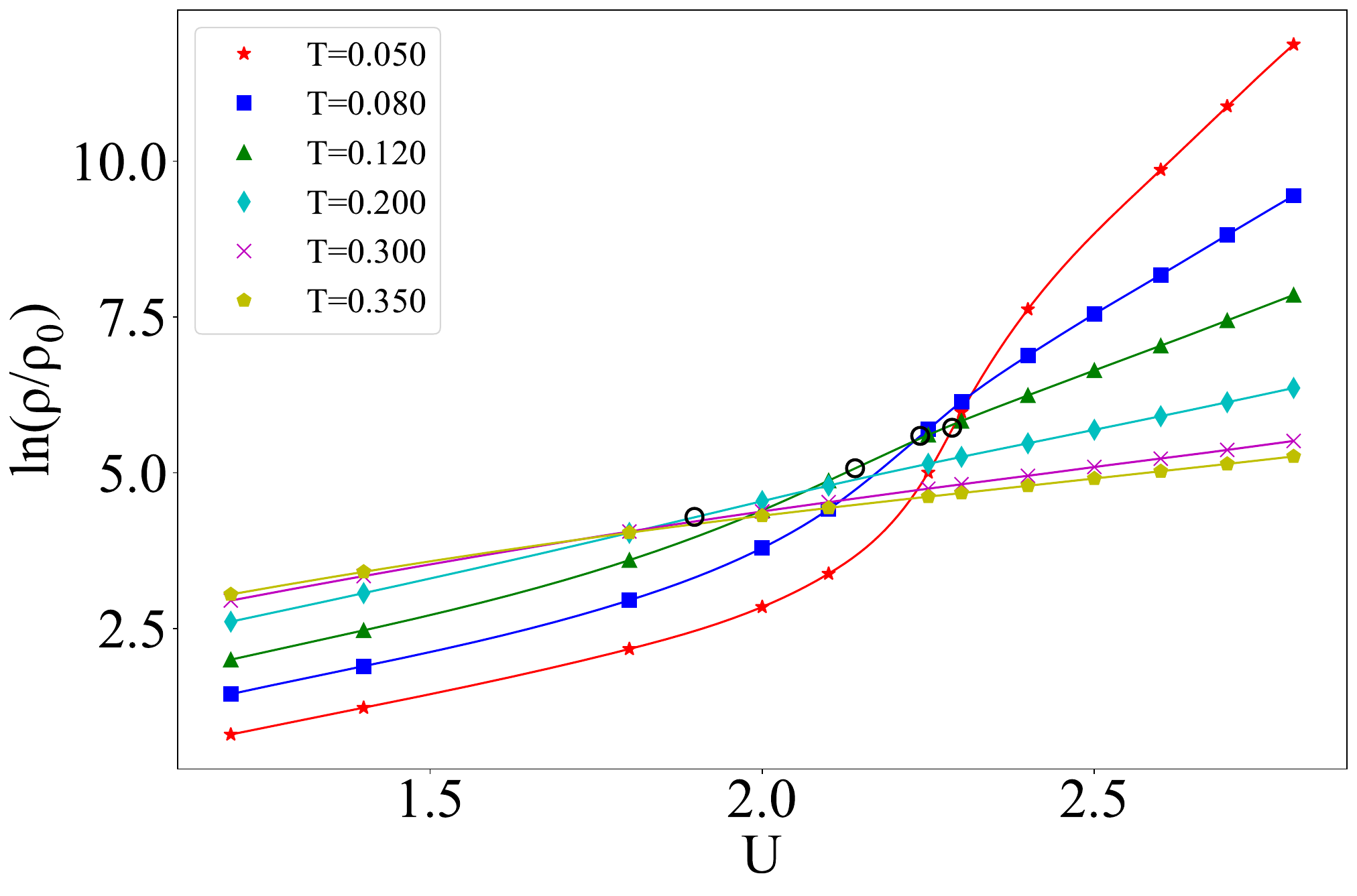}
\caption{(Color online). Interaction dependence of the resistivity in logarithmic scale. The open circles indicate inflection points of the $\ln \rho (U)$.}
\label{fig:resistivity_U}
\end{figure} 

\begin{figure*}[t]
\includegraphics[width=0.7\linewidth]{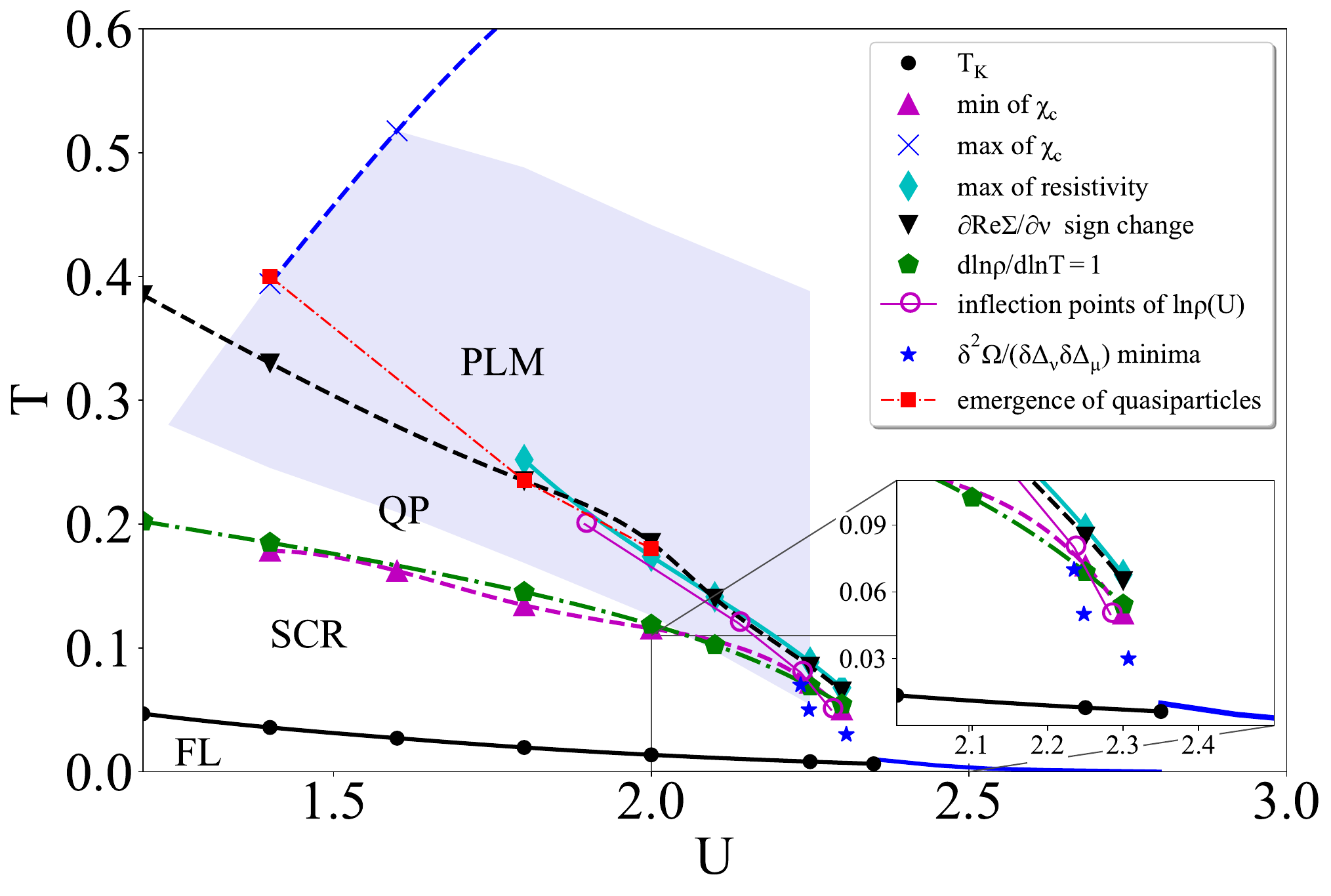}
\caption{(Color online). The obtained phase diagram at $n=1$. The turquoise line (rhombs) shows maxima of $\rho(T)$ and the green dash dotted line (pentagons) indicates points corresponding to the exponent $\beta\equiv d\ln\rho/d\ln T=1$. The temperatures of appearance of quasiparticles obtained from the change of the sign of $\partial {\rm Re} \Sigma(\nu)/\partial\nu$ are shown by black dashed line (triangles) and those obtained from appearance of the peak of the spectral function are shown by the red dashed line (squares). The Widom line determined from the inflection points of the $\ln \rho (U)$ is shown by the purple line (open circles) and that from the minima of the second derivative of the Landau functional $\Omega$ is shown by blue stars. The interaction dependence of the Kondo temperature $T_K$ (black line with circles), the temperatures $T_{c,{\rm max}}$ and $T_{c,{\rm min}}$ of the maxima and minima of local charge compressibility $\chi_c(T)$ (blue dashed line with crosses and purple dashed line with triangles) are taken from  Ref. \cite{OurMott}. The shaded area corresponds to the ``plateau'' of the temperature dependence of effective local moment, see Ref. \cite{OurMott}. 
The critical interaction $U_{c2}$ of the MIT taken from Ref. \cite{Triangular1} is shown by
the blue line, see text for the other notations. The inset zooms the region near {the} MIT.}
\label{fig:phase_diagram_new}
\end{figure*}

The damping of the quasiparticles $\Gamma=-{\rm Im}\Sigma(0)$ at intermediate temperatures increases with increasing $U$ and at the temperatures $T\sim T^*$ follows the linear temperature dependence $\Gamma\simeq \Gamma_0+A T$ instead of the quadratic one for the Fermi liquid. We note that recently similar linear temperature dependencies of the quasiparticle damping were obtained in the combination of ab initio and DMFT studies of vanadium in the regime of partly formed local magnetic moments\cite{OurVanadium}, albeit in the latter case it is caused by Hund interaction instead of proximity to Mott insulator.

To determine the position of the Widom line from the resistivity, we following Refs. \cite{OrgQC,QC2} determine also flection points of the $\ln \rho(U)$ dependencies (see Fig. \ref{fig:resistivity_U}). We find that the flection points coincide to a good accuracy with the above discussed positions of maxima of $\rho(T)$ dependencies, showing the uniquness of determining of the crossover from metaiic to insulating phase near Mott transition. Notably, the obtained dependencies $\ln \rho(U)$ remind closely the experimental data of Ref. \cite{OrgQC} of the pressure dependence $\ln \rho(p)$ in organic layered compound $\kappa$-(ET)$_2$Cu$_2$(CN)$_3$.




The obtained results for the dependence of characteristic temperatures on the interaction strength are combined in Fig. \ref{fig:phase_diagram_new} with previous results of Ref. \cite{OurMott} for characteristic temperatures, obtained from the local charge and spin susceptibilities. We use the notations introduced in Ref. \cite{OurMott}: PLM stands for the preformed local moment regime, SCR is the regime of the local moment screening, and FL denotes the Fermi liquid state.
As mentioned above, the temperatures $T^*$ of the maxima of resistivity and appearance of quasiparticles in both the self-energy and local spectral functions almost coincide, and mark the boundary of the quasiparticle (metallic) regime, which we denote by QP. The Widom line obtained from inflection points of $\ln \rho(U)$ appears to be also close to $T^*(U)$. We also note that the minima of the second derivative of the Landau functional are sufficiently close to the considered boundary, although do not coincide with it precisely. 

Importantly, the temperature $T^*(U)$ of appearance of quasiparticles, being close to the boundary of the plateau of the square of the effective local moment $T\chi_s(T)$ ($\chi_s$ is the local spin susceptibility, see Ref. \cite{OurMott}), is larger than the previously determined temperatures $T_{c,{\rm min}}$ of the minima of local charge compressibility, which mark the boundary of screening of local magnetic moments \cite{OurMott}. This shows that quasiparticles appearing at $T^*(U)$, start to screen local magnetic moments at lower temperatures closer to $T_{c,{\rm min}}$, where they become well defined. To see the trace of the screening of local moments on the temperature dependence of resistivity, we also show on the phase diagram the line of the exponents of resistivity $\beta=d\ln\rho/d\ln T=1$, which is rather close to the $T_{c,{\rm min}}$ boundary. Therefore, start of the screening of local magnetic moments is associated with linear temperature dependence of resistivity, which is inherited from the linear temperature dependence of the self-energy. This gives a possibility of experimental determination of the boundary of the beginning of the screening local magnetic moments.

\subsection{Away from half filling, \texorpdfstring{$n<1$}{Lg}}

\begin{figure}[t]
\includegraphics[width=1.0\linewidth]{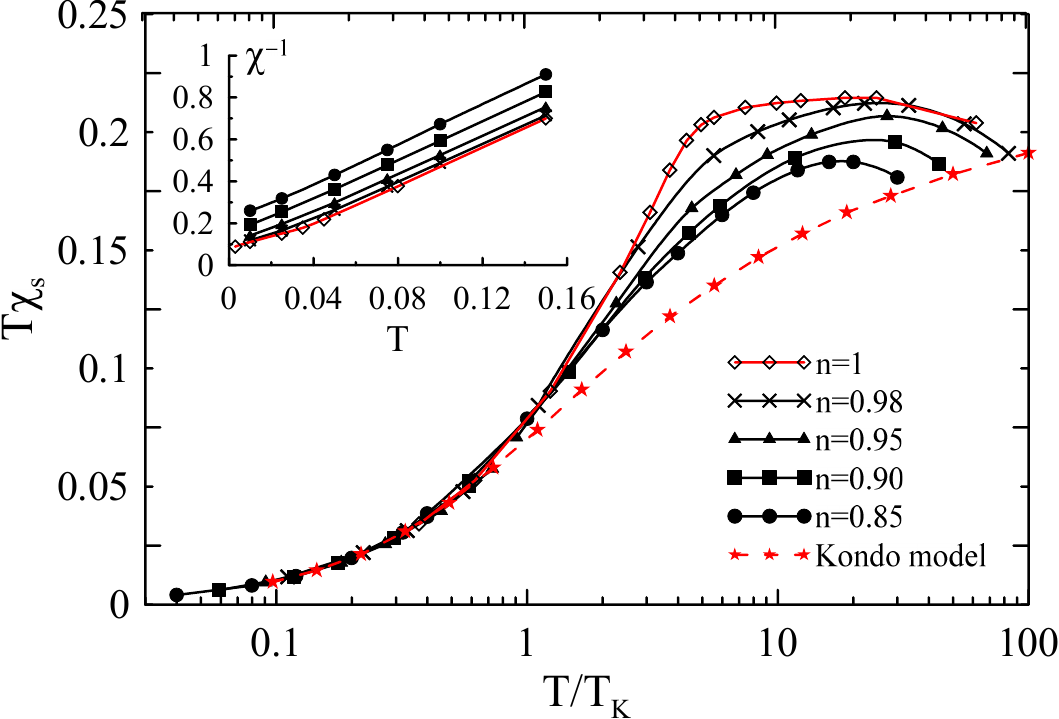}
\caption{(Color online). Temperature dependence of the square of the effective local magnetic moment $\mu_{\rm eff}^2=T\chi_s$ for different values of filling $n$ and $U=2.25$. The open rhombs correspond to $n=1$, crosses to $n=0.98$, triangles to $n=0.95$, squares to $n=0.90$, and circles to $n=0.85$. Dashed line with stars shows the universal temperature dependence for the Kondo model \cite{Wilson,Wilson1}. The inset shows temperature dependence of the inverse local spin susceptibilities.}
\label{fig:chisloc}
\end{figure}

We first investigate the existence and screening of local magnetic moments away from half filling similarly to the way it was done at half filling in Ref. \cite{OurMott} at half filling. In Fig. \ref{fig:chisloc} we show the temperature dependence of the square of the effective local magnetic moment $\mu_{\rm eff}^2=T\chi_s$ and the inverse static local spin susceptibility $\chi_s^{-1}$, where $\chi_s=\int_0^\beta \langle s^z(\tau) s^z(0)\rangle$, $s^z(\tau)$ is the local spin operator in the Heisenberg representation at the imaginary time $\tau$. The inverse local susceptibility is almost linear in temperature, which shows existence of LMM in the low-temperature phase. Similarly to the half filling \cite{OurMott}, we determine the Kondo temperatures $T_K$ by fitting the obtained temperature dependenies of $\mu_{\rm eff}(T)$ to the universal temperature dependence for the Kondo model \cite{Wilson,Wilson1}. One can see that the doping yields reduction of the maximal effective local magnetic moment, which is reached at $T\sim 10 T_K$. The plateau of the temperature dependence of $\mu_{\rm eff}^2$, which was obtained in the half filled case \cite{OurMott}, continuously disappears with doping, such that the Curie law $\chi_{\rm loc}\propto 1/T$ is not fulfilled in the intermediate temperature range. At the same time, as it is mentioned above (see also the inset of Fig. \ref{fig:chisloc}), in a broad range of temperatures the inverse local spin susceptibility remains linear in tempearture, fulfilling the Curie-Weiss law $\chi_{\rm loc}\propto 1/(T+T_{\rm W})$ with the Weiss temperature $T_W \approx \sqrt{2} T_K$, as suggested for the screened regime of the Kondo model \cite{Wilson,Wilson1,Melnikov,Tsvelik}. In contrast to the half filled case, this temperature dependence of the susceptibility is observed 
almost up to the maximum of the dependence $\mu_{\rm eff}(T)$. This shows that at finite doping the LMM exist mainly in the screened regime.

\begin{figure}[t]
\includegraphics[width=1.0\linewidth]{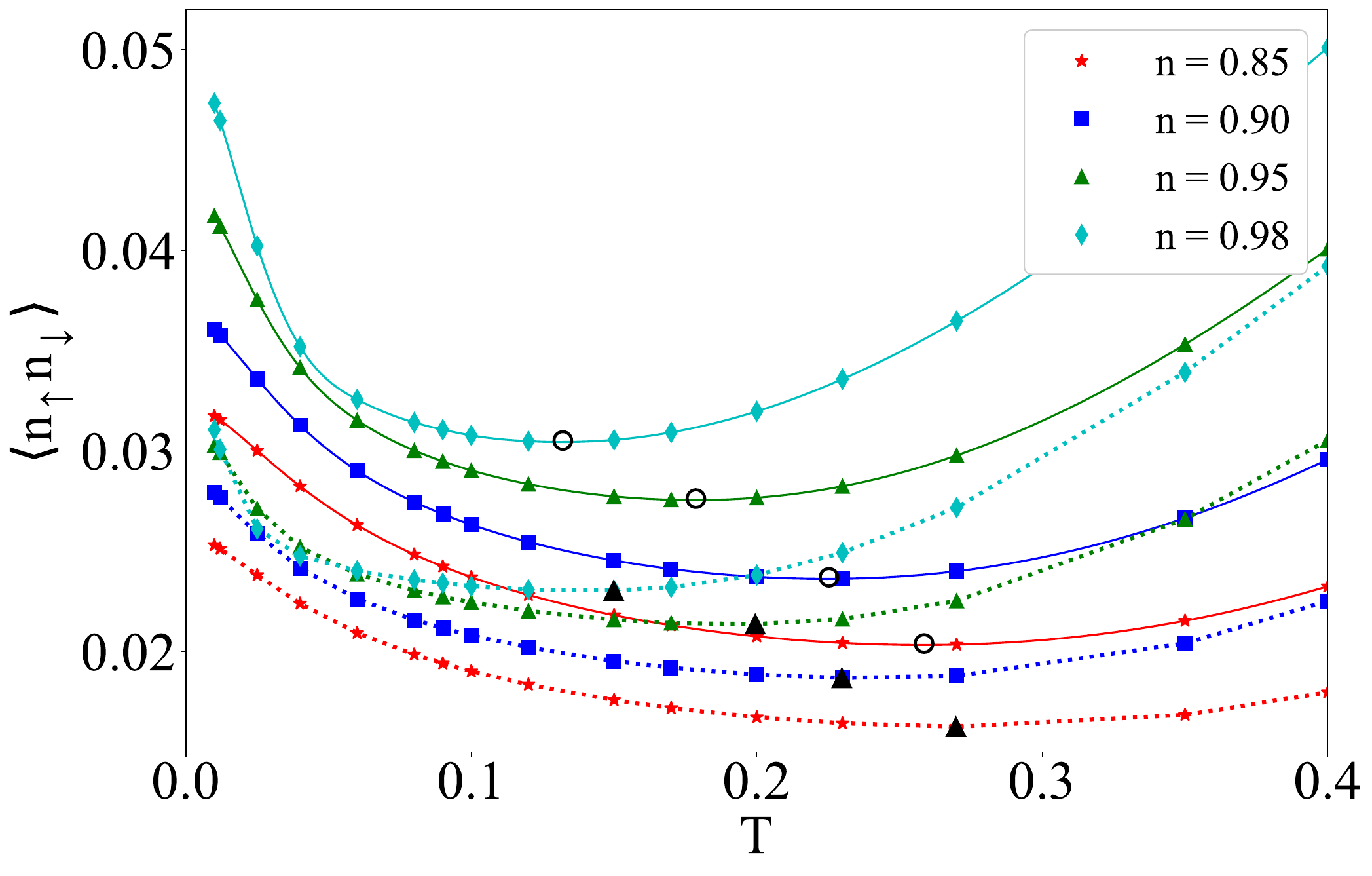}
\caption{(Color online). Temperature dependence of the double occupation for different fillings $n$. Solid lines show data for $U=2.25$, dotted lines are for $U=2.5$. The black open circles (triangles) indicate minima of $\left\langle n_\uparrow n_\downarrow\right\rangle$ for $U=2.25$ ($U=2.5$).}
\label{fig:n_up_n_down_n_2.25}
\end{figure}

\begin{figure}[b]
\includegraphics[width=1.0\linewidth]{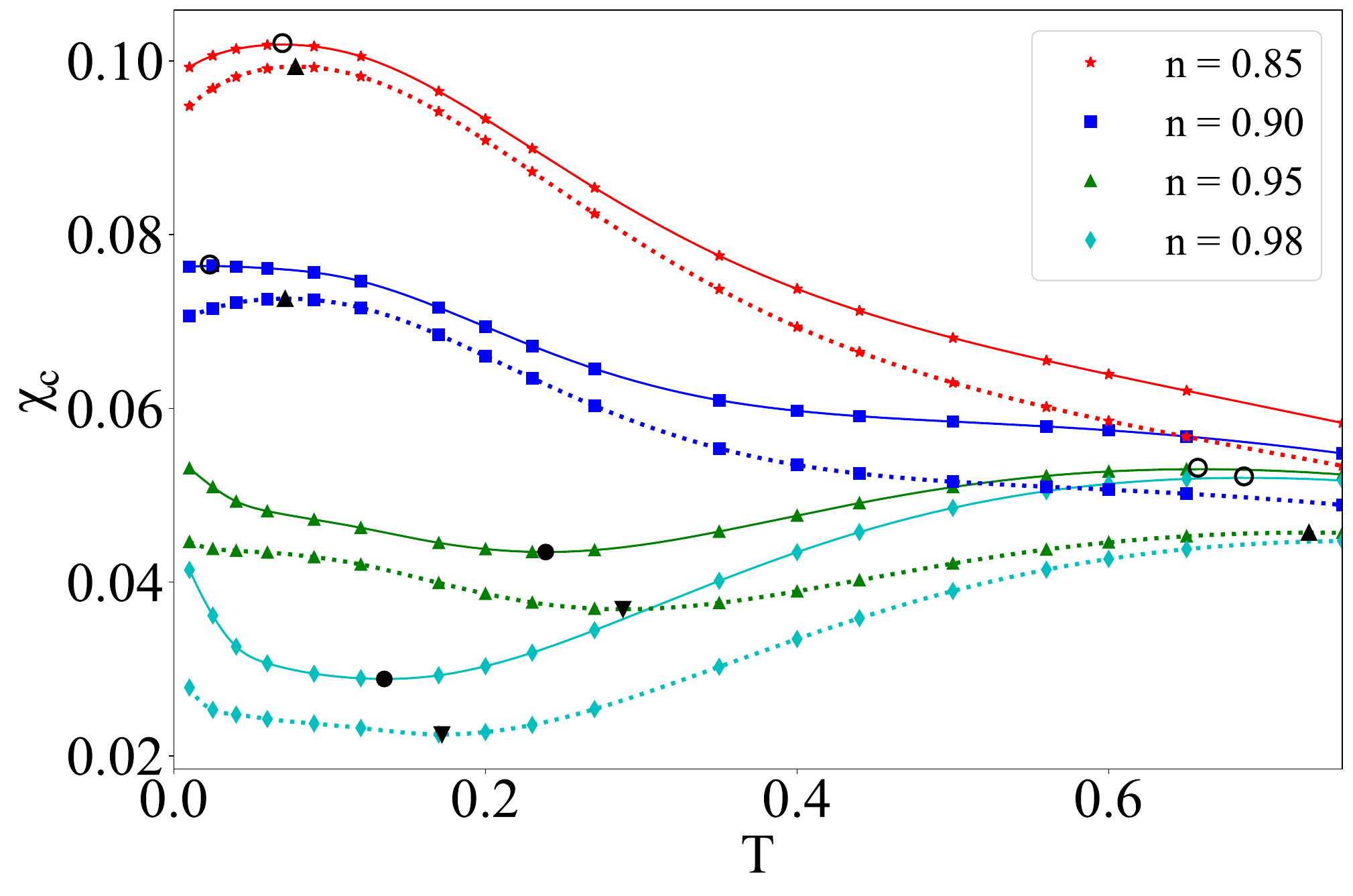}
\caption{(Color online). Temperature dependence of the local static charge susceptibility $\chi_c$ at various fillings $n$. {Solid lines show data for $U=2.25$, dotted lines are for $U=2.5$. The black solid circles (inverted triangles) indicate minima of $\chi_c$ for $U=2.25$ ($U=2.5$), the black opened circles (triangles) indicate maxima of $\chi_c$ for $U=2.25$ ($U=2.5$).}}
\label{fig:chi_c_n_2.25}
\end{figure}

The upper temperature boundary $T_{\rm SCR}$ of the regime of screening of local magnetic moments (which we denote in the following as SCR similarly to half filling),
can be obtained from the minima of the temperature dependence of double occupation (see Fig. \ref{fig:n_up_n_down_n_2.25}). In agreement with the discussion above, the corrresponding temperatures are located somewhat below the maxima of the local magnetic moment $\mu_{\rm eff}(T)$.
To understand peculiarities of SCR regime, we show in Fig. \ref{fig:chi_c_n_2.25} the temperature dependence of local charge compressibilities $\chi_c(T)=dn/d\mu$, where the change of the chemical potential $d\mu$ acts only at the impurity site. For $n=0.98$ and $n=0.95$ we observe the behavior similar to that obtained at half filling \cite{OurMott} with the temperatures of the maxima of local compressibility $T_{c,{\rm max}}$ corresponding to the beginning of formation of local magnetic moments and the temperatures of minima of compressibility $T_{c,\rm{min}}$ corresponding to the   beginning of the screening of LMM. For these fillings, which are close to half filling, the minimum of charge compressibility appears to be close to that of double occupation, such that the two criteria of the screening of LMM agree with each other (cf. Ref. \cite{OurMott}).

\begin{figure}[t]
\includegraphics[width=1.0\linewidth]{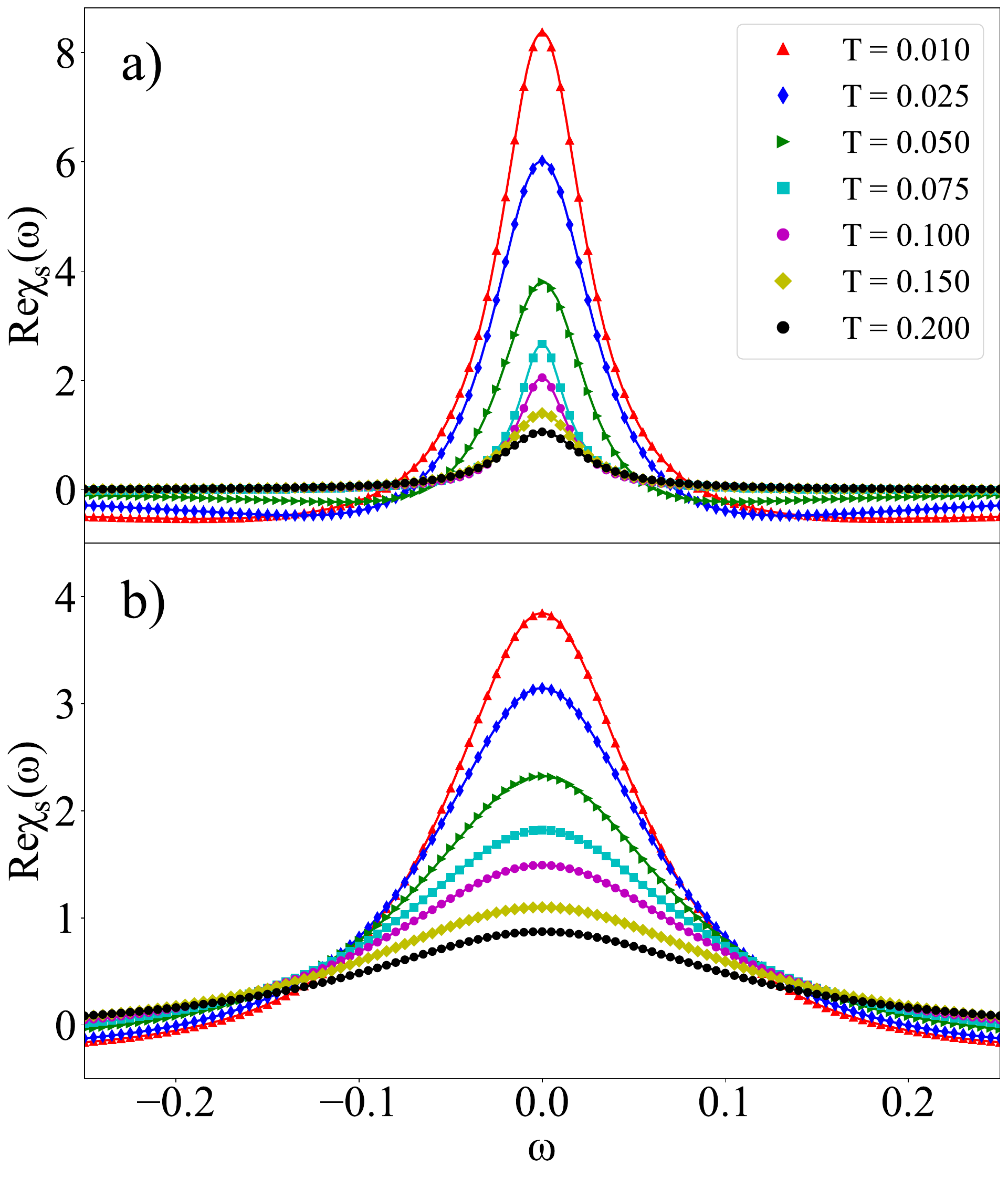}
\caption{(Color online). Real frequency dependence of the real part of the local spin susceptibility $\chi_s(\omega)$ for $U = 2.25$ and $n=0.98$ (a) and $n=0.85$ (b).}
\label{fig:chislocw}
\end{figure}

At the same time, for $n<0.95$ the maximum of the local charge compressibility shifts to much lower temperature $T_{c,{\rm max}}\sim 0.1$ deep inside the SCR regime, and the local minimum of compressibility is not observed. Physically, this appears due to sufficiently large number of free coherent charge carriers. Accordingly, these carriers substantially contribute to the local charge compressibility, changing its temperature dependence, which can be interpreted as the delocalization of hole motion. Nevertheless, the local magnetic moments persist (although in the screened regime) even in this case, which can be seen from the above discussed temperature dependencies of static local spin susceptibility, as well as from the frequency dependencies of the dynamic local spin susceptibilities $\chi_s(\omega)$. Indeed, the dynamic susceptibilities show a sharp peak at $\omega=0$ (see Fig. \ref{fig:chislocw}), which is chracteristic for the presence of LMM \cite{OurFe1, OurFeGamma}. In contrast to the half filled case \cite{OurMott} the width of the peak increases with decrease of temperature, which shows that the local moments lifetime decreases. Also, the width of the peak increases with doping. However, the half width of the peak $\Delta \omega\sim 0.03-0.05$ remains smaller than the temperature of the crossover to the SCR regime, which implies that the local moments in the substantial part of SCR regime remain well defined.

\begin{figure}[t]
\includegraphics[width=1.0\linewidth]{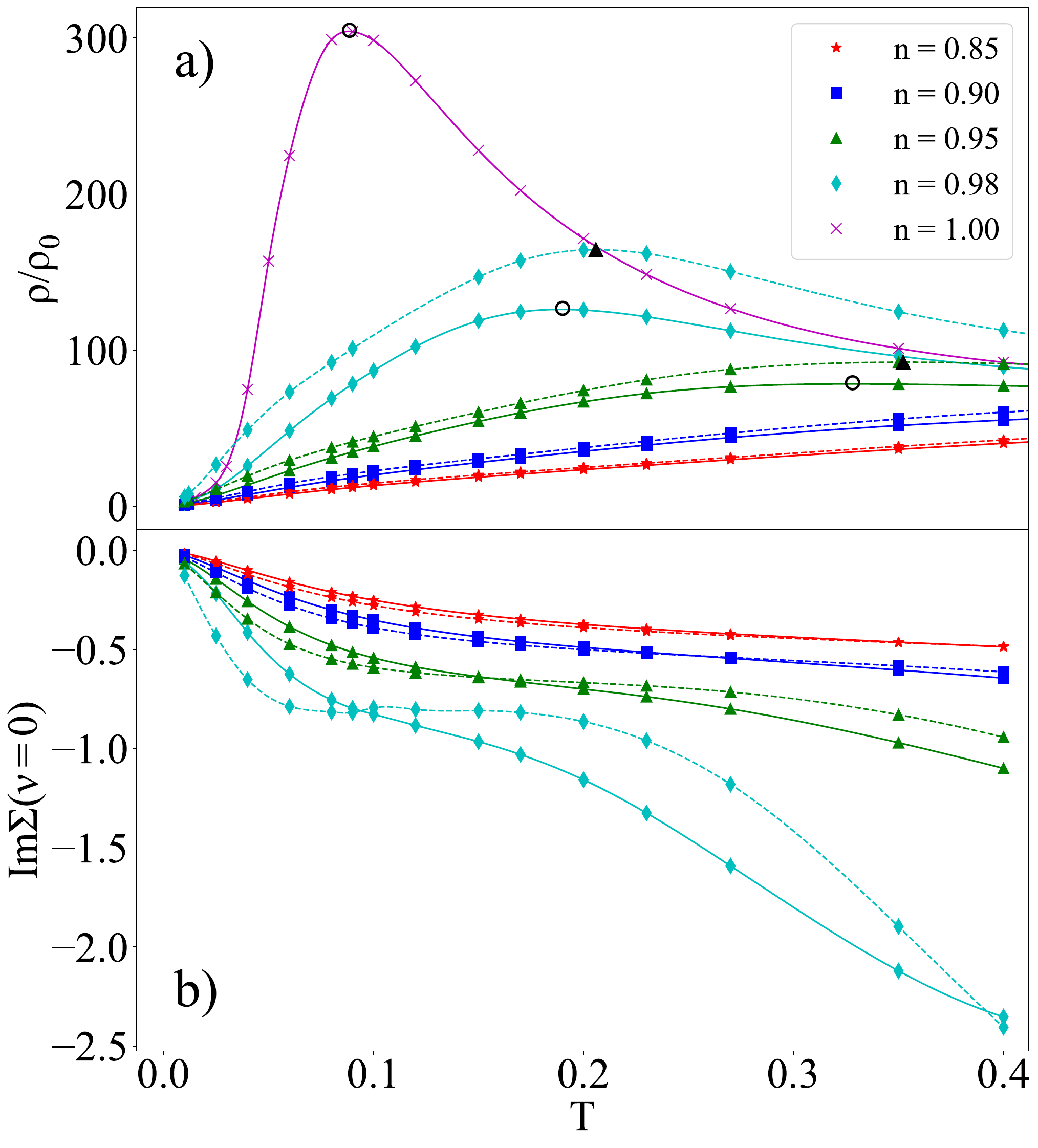}
\caption{(Color online). Temperature dependencies of (a) the resistivity (in units of $\rho_0=\hbar/(4e^2)$), (b) the imaginary part of zero-frequency self-energy for various fillings $n$. Solid lines correspond to $U=2.25$, dotted lines are for $U=2.5$. The black open circles (triangles) indicate maxima of $\rho$ for $U=2.25$ ($U=2.5$). }
\label{fig:resistivity_n1}
\end{figure}

\begin{figure}[t]
\includegraphics[width=1.0\linewidth]{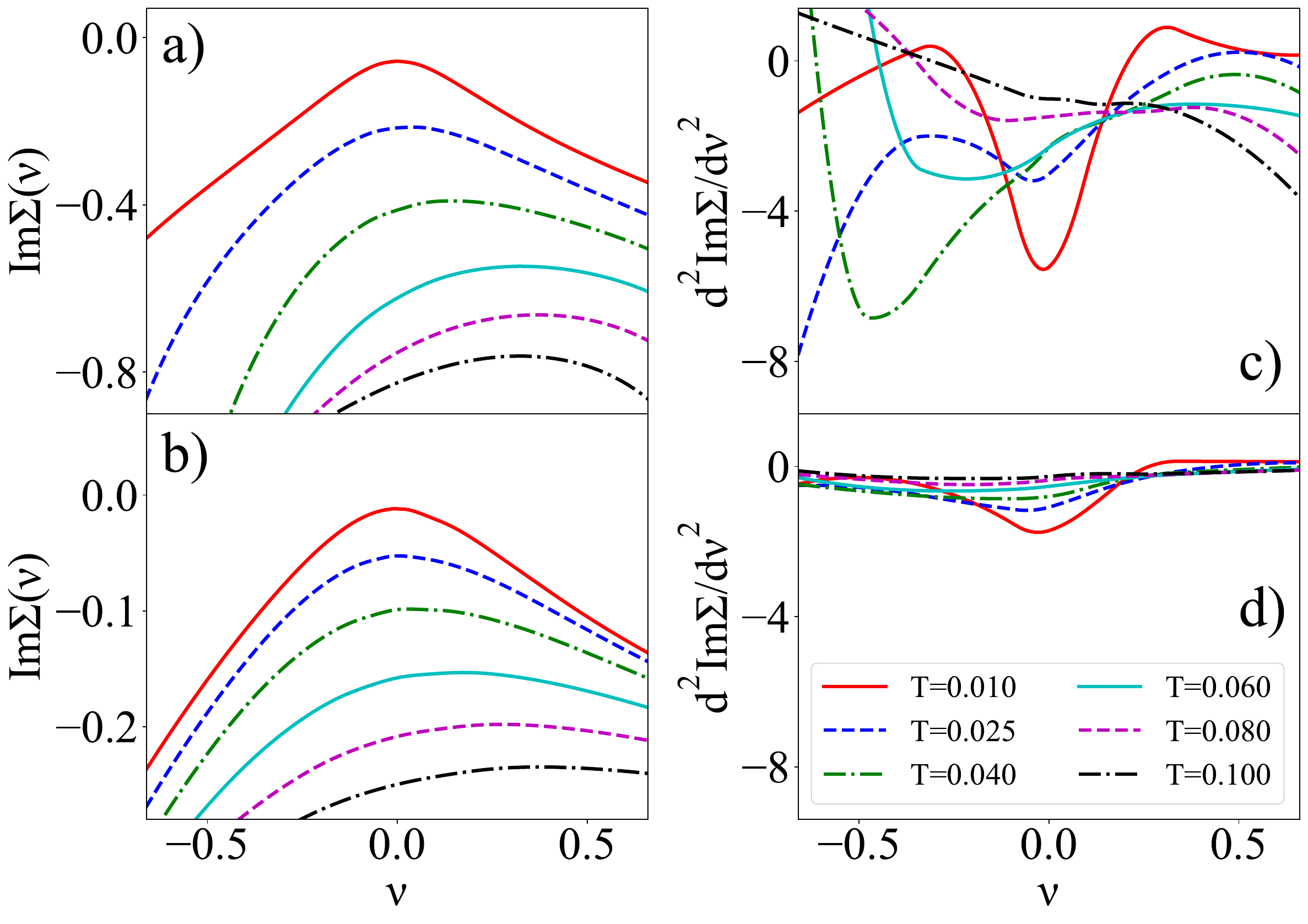}
\caption{(Color online). The frequency dependencies of the imaginary part of the self-energy (a,b) and its second derivative (c,d) for $U = 2.25$ and the filling $n=0.98$ (a,c) and $n=0.85$ (b,d). }
\label{fig:im_sigma4}
\end{figure}

Let us discuss how 
the temperature dependence of the resistivity changes 
with the decrease of the filling $n$. 
In Fig. \ref{fig:resistivity_n1}a we show the temperature evolution of the resistivity for $U=2.25$ (corresponding to metallic phase at half filling) and $U=2.5$ (corresponding to the insulating phase at half filling). As in the case of half-filling and similarly to previous studies  of doped Hubbard model on the infinite dimensional hypercubic \cite{Mott_res1a} and Bethe \cite{QC3} lattice, close to half filling there is a maximum of the resistivity at a certain temperature $T^*(U, n)$, which increases with decrease of $n$.
Suppression of the resistivity in comparison to the half filling $n=1$ is related to smaller damping of the quasiparticles away from half filling (see below). The peak of the resistivity also becomes less pronounced, from which it can be assumed that the boundary between coherent and incoherent quasiparticle regimes becomes even less defined, than for half filling.  At the same time, the region of almost linear behavior of resistivity becomes broader than for $n=1$, as it was obtained previously in Ref. \cite{Mott_res1a} for the infinite dimensional hypercubic lattice and in Ref. \cite{Triangular1} for the square and triangular lattice. 


The linear temperature dependence of the resistivity originates from the temperature dependence of the imaginary part of the self-energy (Fig. \ref{fig:resistivity_n1}b), which shows almost linear behavior $\Gamma\simeq A T$ (with a small zero-frequency temperature-independent part) in the temperature range $T<0.05$ and more complex behavior at higher temperatures; as mentioned above, the quasiparticle damping remains smaller than for $n=1$ and decreases with decrease of filling. The frequency dependence of the self-energy is shown in Fig. \ref{fig:im_sigma4}. In contrast to the half filled case, we observe quasiparticle behaviour for all considered temperatures. With decrease of filling the second derivative of the self-energy with respect to frequency also decreases, which implies closer to linear in frequency behavior of the self energy.

\begin{figure}[b]
\includegraphics[width=1.0\linewidth]{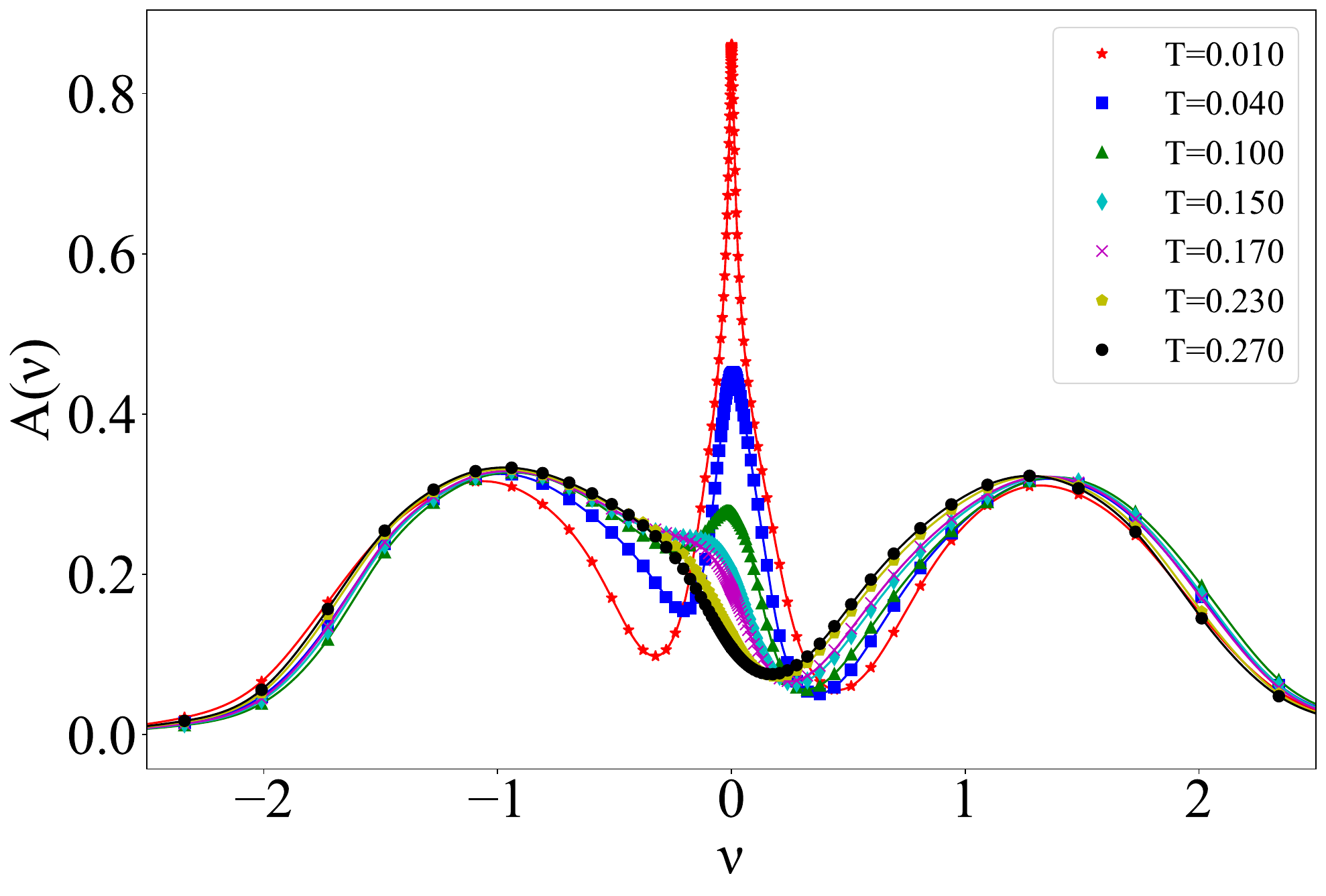}
\caption{(Color online). The frequency dependencies of the spectral function for $U = 2.25$, the filling $n=0.98$, and various temperatures $T$. The temperature range  $T>0.17$ is associated with PLM+H regime, where the central quasiparticle peak is absent and the holes move incoherently in the background of local magnetic moments. The SCR regime ($T<0.1$) is characterised by the narrow quasiparticle peak at the Fermi level.
}
\label{fig:spectral_n_2.25_0.98}
\end{figure}

\begin{figure}[t]
\includegraphics[width=1.0\linewidth]{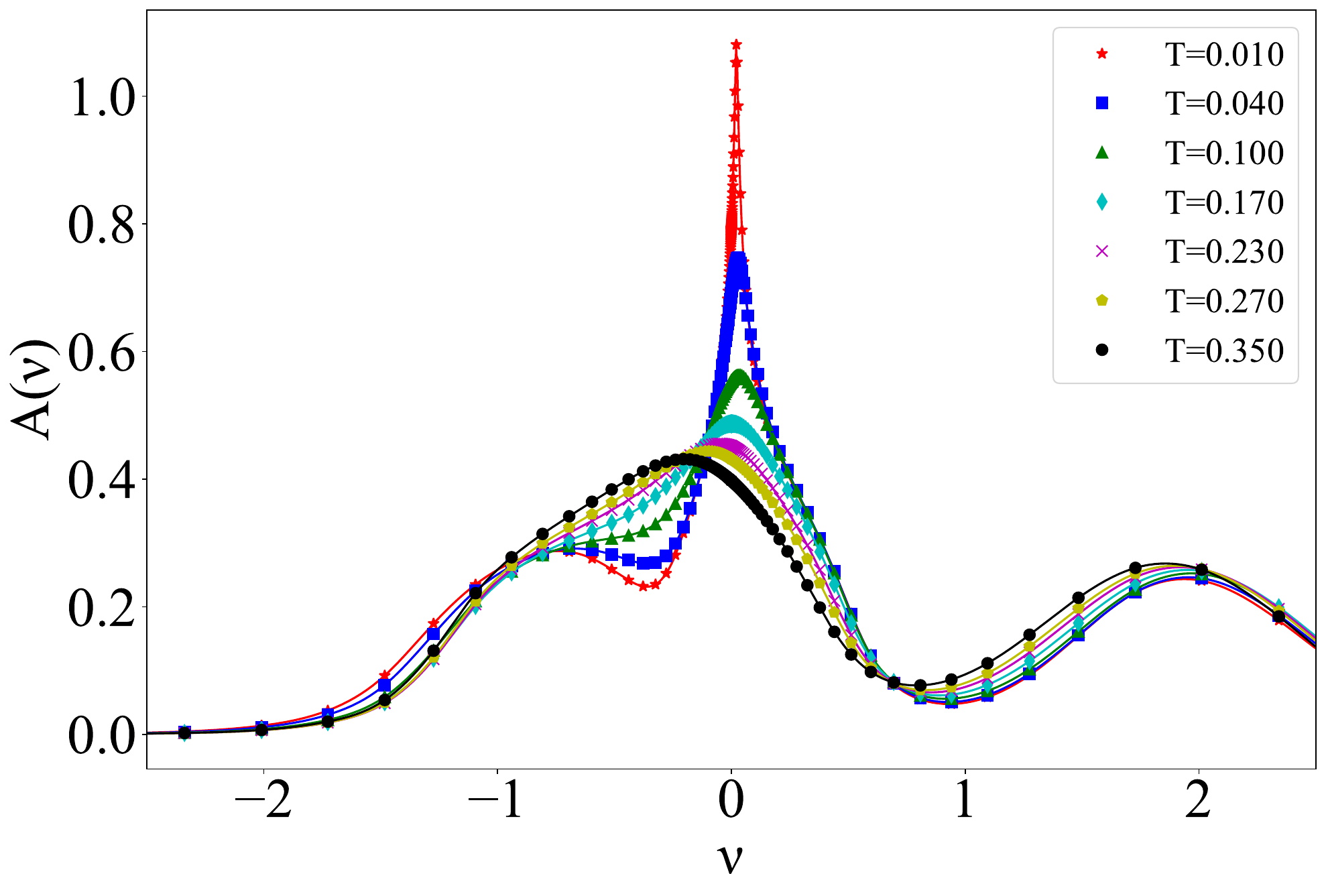}
\caption{(Color online). The same as Fig. \ref{fig:spectral_n_2.25_0.98} for $n=0.85$. The SCR regime ($T<0.23$) is characterised by the narrow quasiparticle peak at the Fermi level.
}
\label{fig:spectral_n_2.25_0.85}
\end{figure}

Similarly to half filling, the temperatures of the maxima $T^*(U,n)$ are located somewhat above the temperatures of the onset of screening  $T_{\rm SCR}$. To emphasize the origin of the maximum of the temperature dependence of resistivity at $n<1$, in Fig. \ref{fig:spectral_n_2.25_0.98} we show temperature evolution of the spectral functions for $n=0.98$. One can see that the form of the spectral functions changes with increase of temperature from having central quasiparticle peak at the Fermi level at low temperatures, similar to that observed at half filling \cite{OurMott}, to the monotonously decreasing density of states near the Fermi level, located at the upper edge of lower Hubbard band. This behavior is similar to that for the particle-hole asymmetric Anderson impurity model  \cite{Horvatic,Hewson}.
The high-temperature form of the spectral functions (refered below as PLM+H regime) physically corresponds to holes, which move incoherently in the background of local magnetic moments. At the same time, the quasiparticle peak at low temperatures corresponds to the many-particle screening state. Near half filling (i.e. for $n=0.98$) the temperature of the appearance of the quasiparticle peak, determined by appearance of the inflection points of the energy dependence of density of states near the Fermi level, is close to the temperature $T^*$ at which maximum of the resistivity is observed; the temperature $T^*$ also coincides with reaching maximal value of the local magnetic moment (cf. Fig. \ref{fig:chisloc}). Therefore, in this regime we associate the maximum of resistivity with the crossover from the incoherent motion of holes to the presence of coherent quasiparticles.

Importantly, the crossover with appearing the quasiparticle peak at the Fermi level
with lowering temperature 
occurs at all considered fillings $n\geq 0.85$, while the Fermi level shifts from the upper edge to the center of the lower Hubbard subband with increase of the doping to $\delta=1-n\sim 0.15$ (see Fig. \ref{fig:spectral_n_2.25_0.85}). In the latter case the Fermi level is far below the interaction-induced minimum of the local density of states, and, as it is discussed above, the LMM exist only in the screened state at low temperatures.
On the other hand, the high temperature form of the spectral function corresponds to coherently moving holes,
refered below as the CMH regime. The transition between PLM+H and CMH regimes corresponds to changing temperature dependence of the local charge susceptibility, discussed above for Fig. \ref{fig:chi_c_n_2.25}.

As mentioned above, at low temperatures 
almost linear temperature dependence of the quasiparticle damping and resistivity occurs, which is reminiscent of the marginal Fermi liquid of Ref. \cite{Varma}. We can therefore interpret this state as occurring due to 
competition of two effects: increase of the coherence of quasiparticles due to the presence of sufficient number of charge carriers and their decoherence due to interaction with local magnetic moments. We note that similar linear temperature dependence of resistivity was previously obtained in the doped Kondo lattice model \cite{Lavagna} and compactified version of the single-impurity Anderson model \cite{CompAnd}.

\begin{figure}[t]
\includegraphics[width=1.0\linewidth]{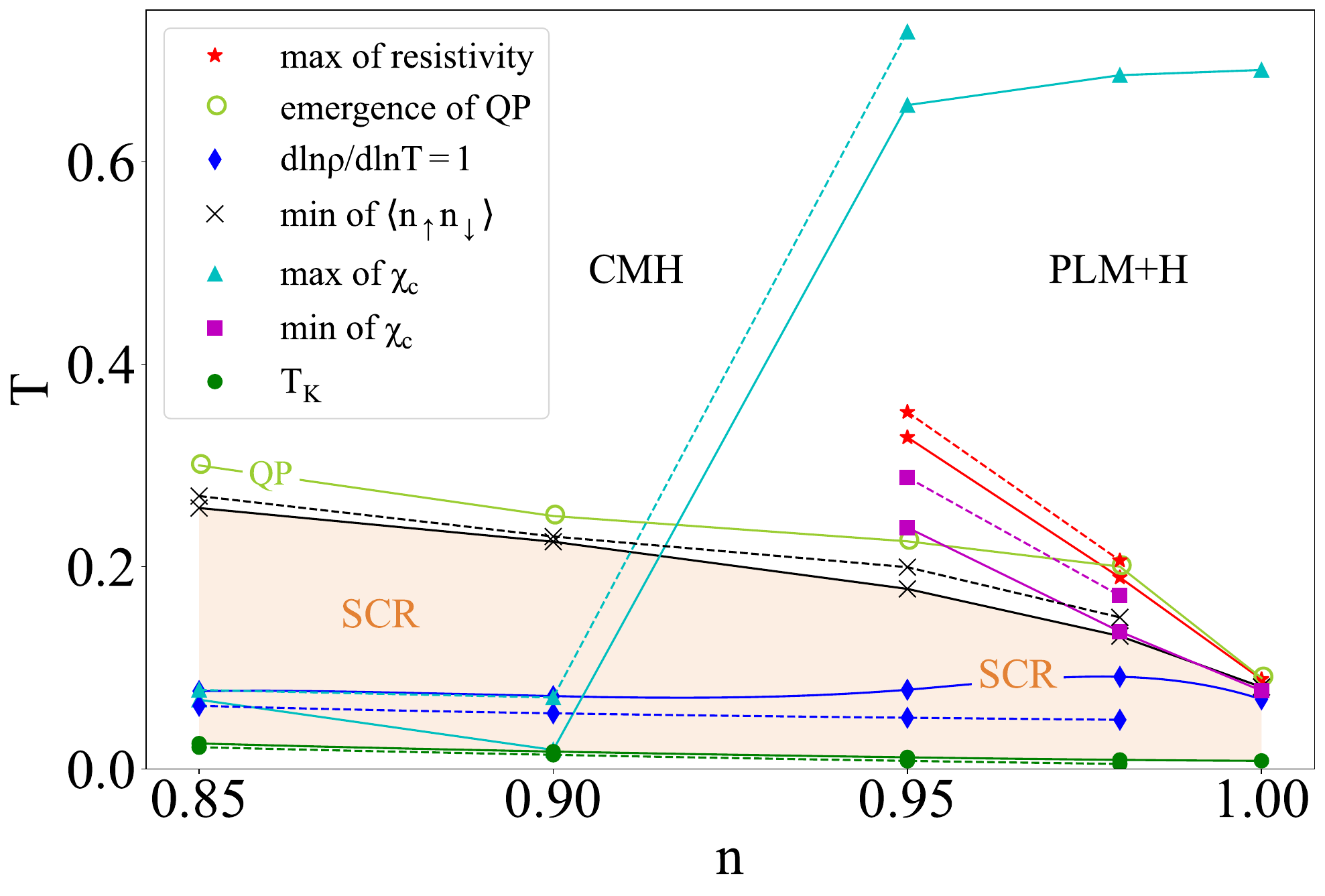}
\caption{The phase diagram away from half filling. Solid lines show data for $U=2.25$, dotted lines are for $U=2.5$. The turquoise line (triangles) shows maxima of $\chi_c(T)$, corresponding to the upper temperature boundary of the unscreened LMM, and the purple line (squares) corresponds to minima of $\chi_c(T)$, which mark beginning of screening of LMM in PLM+H regime. The green line with open circles shows the appearance of sharp quasiparticle peak (corresponding to the crossover to the QP regime). The black lines with crosses shows minima of double occupation $\left\langle n_\uparrow n_\downarrow\right\rangle$, which bounds the screening SCR regime (shaded orange area for $U=2.25$); the Kondo temperatures are shown by green lines (circles). The red lines (stars) indicate maxima of $\rho(T)$, corresponding to the crossover from incoherent to coherent quasiparticle motion, and blue line (rhombs) shows $\beta\equiv d\ln\rho/d\ln T=1$ points, which mark the onset of the linear temperature dependence of quasiparticle damping and resistivity.   
For the form of the spectral functions in different regimes see Figs. \ref{fig:spectral_n_2.25_0.98}, 
\ref{fig:spectral_n_2.25_0.85}.
}
\label{fig:resistivity_features}
\end{figure}

The obtained results for $n<1$ are collected in the phase diagram of Fig. \ref{fig:resistivity_features}. Close to half filling the temperature of the maximum of the resistivity, which determines the boundary of PLM+H regime and the regime with present coherent quasiparticles, is somewhat larger than the temperatures of the minimum of double occupation and charge compressibility, which mark the onset of the screening regime $T_{\rm SCR}$. As described above, the screening of local magnetic moments in this regime is similar to that at half filling. {With moving further away from half filling the temperature of the onset of quasiparticles approaches the boundary of the SCR regime, which shows again that the local magnetic moments in this regime exist in the screened phase.}
In the interval of fillings $0.90 < n < 0.95$ the temperature of the local compressibility maximum sharply changes towards low temperatures, which, as we noted earlier, indicates delocalization of the charge carriers (i.e. holes).
This idea is also supported by the fact that far away from half filling the resistivity $\rho(T)$ becomes monotonously increasing with temperature, which shows different nature of fermionic quasiparticles close and far away from half filling, corresponding to the crossover from PLM+H to CMH regime. 

The upper temperature boundary of the SCR regime, determined from the minimum of double occupation, increases only moderately with decrease of $n$. As mentioned above, close to half filling this boundary is close to $T_{c,{\rm min}}$. 
At low temperatures $T\lesssim T_{\beta=1}$, where $T_{\beta=1}\sim 0.1$ is the temperature, at which the exponent $\beta\equiv d\ln\rho/d\ln T=1$, we find almost linear behavior of the quasiparticle damping and the resistivity, which, as mentioned above, reminds the marginal Fermi liquid state.  With increase of the doping the linear behavior of resistivity becomes more pronounced.  

\section{Conclusions}

In the present paper we have considered the effect of local magnetic moments on the temperature dependence of resistivity. We have shown that the maximum of the temperature dependence of resistivity, obtained with DMFT, $T^*(U,n)$ corresponds to beginning of the formation of quasiparticle peak of the spectral function. Physically the temperature $T^*(U,1)$ marks the crossover from insulating to the coherent quasiparticle behavior at half filling, while away from half filling $T^*(U,n)$ marks the crossover from incoherent to coherent motion of charge carriers and it is obtained only sufficiently close to half filling ($n\geq 0.95$).

{At half-filling, we considered the boundary for the appearance of quasiparticles, which in this case coincides with {with the temperature of maximum of resistivity} and also {close to} the Widom line. An important {aspect is} that {appearance of} quasiparticles does not cause {necessarily} screening of LMM, {the latter} begins only after the {coherence of quasiparticles} 
becomes sufficient with decreasing temperature. The boundary of the beginning of the screening regime, previously defined by the minima of the local charge susceptibility and double occupation, perfectly corresponds to the exponent of the resistivity $\beta\equiv d\ln\rho/d\ln T=1$.}


Close to half filling we find that at not too low temperatures the local magnetic moments exist, together with incoherently moving holes. This is the regime described by the $t-J$ model \cite{Castellani,Hirsch,Cyrot,HighTc}; the present study shows however that it is restricted to few percent doping only. With lowering temperature, the LMM are screened by itinerant electrons, similarly to half filling. With moving further away from half filling 
the hole motion becomes more coherent. In this case the LMM exist only in screened phase at sufficiently low temperatures, 
where the sharp quasiparticle peak of the spectral function emerges.
Presence of both, LMM and coherent quasiparticles at sufficient doping yields at low temperatures linear dependent scattering rate and resistivity below the temperature, at which the exponent  
$\beta=1$ is reached. 

We therefore emphasize the complexity of the obtained physical properties of the doped phase, which on one hand show features of conventional quasiparticle behavior at sufficient doping, but on the other hand shows traces of screened LMM. While at high temperatures the crossover from the incoherent to coherent mortion of holes is obtained, at low temperatures we find SCR phase, which properties are distinctly different from the coherent hole motion. This phase shows linear temperature dependence of the scattering rate and resistivity, reminiscent of the marginal Fermi liquid \cite{Varma}. 
Appearance of incoherent holes at low doping, onset of their coherence at larger doping $\delta\sim 0.1$, and the ``marginal" Fermi liquid behavior at low temperatures highlight some similarities wit the physics of high-temperature superconductors (see, e.g., Ref. \cite{HighTc}). For the description of these compounds, however, magnetic and, possibly, charge correlations are likely have to be taken into account.

Inclusion of magnetic and/or charge correlations, e.g. within the {dynamical cluster approximation (DCA) or cellular dynamical mean-field theory (CDMFT) approach} and/or nonlocal diagrammatic extensions of DMFT \cite{OurRev} is therefore of certain interest for future research. Another interesting topic is studying the effect of Hund exchange in multiorbital models and considering the difference of ``Mottness" and ``Hundness" behavior of local magnetic moments.

{\it Note added}. After finishing the paper we have learned about the related study of the half filled Hubbard model on a square lattice \cite{Yang}, which confirms coincidence of  the Widom line $T_{\rm QWL}(U)$ with the temperatures of appearance of quasiparticles, as well as maximum of resistivity, and the screening of local magnetic moments at the lower temperatures, $T\le T_{\rm SCR}<T_{\rm QWL}$.



\section*{Acknowledgements}

The authors acknowledge the financial support from the BASIS foundation (Grant No. 21-1-1-9-1) and the Ministry of Science and Higher Education of the Russian Federation (Agreement No. 075-15-2021-606). A. A. K. also acknowledges the financial support within the theme ``Quant" AAAA-A18-118020190095-4 of Ministry of Science and Higher Education of the Russian Federation.


\begin{thebibliography}{1}

\bibitem{Mott} N. F. Mott, Proc. Phys. Soc. A {\bf 62}, 416 (1949).

\bibitem{V2O3-1} D. B. McWhan, T. M. Rice, and J. P. Remeika, Phys. Rev. Lett. {\bf 23}, 1384 (1969); D. B. McWhan and T. M. Rice, Phys. Rev. Lett. {\bf 22}, 887 (1969); D. B. McWhan, J. P. Remeika, J. P. Maita,
H. Okinaka, K. Kosuge, and S. Kachi, Phys. Rev. B {\bf 7}, 326 (1973).

\bibitem{V2O3-2} M. Rubinstein, Phys. Rev. B {\bf 2}, 4731 (1970).

\bibitem{V2O3-3} D. B. McWhan, A. Menth, J. P. Remeika, W. F. Brinkman, and T. M. Rice, Phys. Rev. B {\bf 7}, 1920 (1973).

\bibitem{V2O3-4} K. Held, G. Keller, V. Eyert, D. Vollhardt, and V. I. Anisimov, Phys. Rev. Lett. {\bf 86}, 5345 (2001); G. Keller, K. Held, V. Eyert, D. Vollhardt, and V. I. Anisimov, Phys. Rev. B {\bf 70}, 205116 (2004); {D. Vollhardt, K. Held, G. Keller, R. Bulla, Th. Pruschke, I. A. Nekrasov, and V. I. Anisimov, J. Phys.
Soc. Jpn. {\bf 74}, 136 (2005).}

\bibitem{V2O3-5} V. I. Anisimov, D. E. Kondakov, A. V. Kozhevnikov, I. A. Nekrasov, Z. V. Pchelkina, J. W. Allen, S.-K. Mo, H.-D. Kim, P. Metcalf, S. Suga, A. Sekiyama, G. Keller, I. Leonov, X. Ren, and D. Vollhardt, Phys. Rev. B {\bf 71}, 125119 (2005).

\bibitem{V2O3-6} A. I. Poteryaev, J. M. Tomczak, S. Biermann, A. Georges, A. I. Lichtenstein, A. N. Rubtsov, T. Saha-Dasgupta, and O. K. Andersen, Phys. Rev. B {\bf 76}, 085127 (2007).

\bibitem{V2O3-7} P. Hansmann, A. Toschi, G. Sangiovanni, T. Saha-Dasgupta, S. Lupi, M. Marsi, and K. Held, Phys. Status Solidi B {\bf 250}, 1251 (2013).
\bibitem{Org2} Y. Kurosaki, Y. Shimizu, K. Miyagawa, K. Kanoda, and G. Saito, Phys. Rev. Lett. {\bf 95}, 177001 (2005).
\bibitem{OrgQC} T. Furukawa, K. Miyagawa, H. Taniguchi, R. Kato, and K. Kanoda, Nat. Phys. {\bf 11}, 221 (2015).
\bibitem{Org5} 
A. Pustogow, M. Bories, A. L\"ohle, R. R\"osslhuber, E. Zhukova, B. Gorshunov, S. Tomi\'c, J. A. Schlueter, R. H\"ubner, T. Hiramatsu, Y. Yoshida, G. Saito, R. Kato, T.-H. Lee, V. Dobrosavljevi\'c, S. Fratini, and M. Dressel, Nat. Mater. {\bf 17}, 773 (2018). 
\bibitem{Org1} W. Li, A. Pustogow, R. Kato, and M. Dressel, Phys. Rev. B {\bf 99}, 115137 (2019).
\bibitem{OrgQP} A. Pustogow, Y. Saito, A. L\"ohle, M. S. Alonso, A. Kawamoto, V. Dobrosavljevi\'c, M. Dressel, and S. Fratini, Nat. Commun. {\bf 12}, 1571 (2021).
\bibitem{Org3} Y. Saito, A. Löhle, A. Kawamoto, A. Pustogow, and M. Dressel, Crystals {\bf 11}, 817 (2021).
\bibitem{Org4} {A. Pustogow, R. R\"osslhuber, Y. Tan, E. Uykur, A. B\"ohme, M. Wenzel, Y. Saito, A. L\"ohle, R. H\"ubner, A. Kawamoto, J. A. Schlueter, V. Dobrosavljevi\'c, and M. Dressel,}
npj Quantum Materials {\bf 6}, 9 (2021).

\bibitem{HighTc} P. A. Lee, N. Nagaosa, and X.-G. Wen, Rev. Mod. Phys. {\bf 78}, 17 (2006).

\bibitem{Spalek} J. Spalek, Journ. Sol. State Chem. {\bf 88}, 70 (1990).
\bibitem{Nozieres} Ph. Nozi\'eres, J. Phys. Soc. Jpn. {\bf 74}, 4 (2005).
\bibitem{Fabrizio} M. Fabrizio, in \textit{The Physics of Correlated Insulators, Metals, and Superconductors}, edited by E. Pavarini, E. Koch, R. Scalettar, and R. M. Martin, Lecture Notes of the Autumn School on Correlated Electrons, Modeling and Simulation Vol. {7}, {Chapter 13} (Verlag des Forschungszentrum Jülich, 2017).


\bibitem{DMFT} A. Georges, G. Kotliar, W. Krauth, and M. Rozenberg, 
Rev. Mod. Phys. {\bf 68}, 13 (1996); G. Kotliar and D. Vollhardt, Physics Today {\bf 57} (3), 53 (2004).
\bibitem{Mott_res1} Th. Pruschke and D. L. Cox, and M. Jarrell, Phys. Rev. B {\bf 47}, 3553 (1993).

\bibitem{Mott_res1a} M. Jarrell and  Th. Pruschke, Phys. Rev. B {\bf 49} 1458 (1994); Th. Pruschke,M. Jarrell, and J.K. Freericks, Adv. Phys. {\bf 44}, 187 (1995).

\bibitem{Mott_res2} J. Merino and R. H. McKenzie, Phys. Rev. B {\bf 61}, 7996 (2000).

\bibitem{Mott_res3} M. M. Radonji\'c,  D. Tanaskovi\'c, V. Dobrosavljevi\'c, K. Haule, and G. Kotliar, Phys. Rev. B {\bf 85}, 085133 (2012).


\bibitem{QC1} H. Terletska, J. Vu\v{c}i\v{c}evi\'c, D. Tanaskovi\'c, and V. Dobrosavljevi\'c, Phys. Rev. Lett. {\bf 107}, 026401 (2011).

\bibitem{QC2} J. Vu\v{c}i\v{c}evi\'c, H. Terletska, D. Tanaskovi\'c, and V. Dobrosavljevi\'c, Phys. Rev. B {\bf 88}, 075143 (2013).

\bibitem{QC3} J. Vu\v{c}i\v{c}evi\'c, D. Tanaskovi\'c, M. J. Rozenberg, and V. Dobrosavljevi\'c, Phys. Rev. Lett. {\bf 114}, 246402 (2015).

\bibitem{QC4} H. Eisenlohr , S.-S. B. Lee, and M. Vojta, Phys. Rev. B {\bf 100}, 155152 (2019).

\bibitem{Widom1} A. Reymbaut, M. Boulay, L. Fratino, P. S\'emon, W. Wu, G. Sordi, and A. M. S. Tremblay, arXiv:2004.02302.


\bibitem{Widom} B. Widom, in Phase Transitions and Critical Phenomena, (eds C. Domb and M. S. Green) {(Academic Press, New York, 1972), Vol. 2, Chapter 3.}


\bibitem{WidomLG} G. G. Simeoni, T. Bryk, F. A. Gorelli, M. Krisch, G. Ruocco, M. Santoro, and T. Scopigno, Nat. Phys. {\bf 6}, 503 (2010).

\bibitem{Toschi} P. Chalupa, T. Sch\"afer, M. Reitner, D. Springer, S. Andergassen, and A. Toschi, 
Phys. Rev. Lett. {\bf 126}, 056403 (2021).
\bibitem{Katsnelson} E. A. Stepanov, S. Brener, V. Harkov, M. I. Katsnelson, and A. I. Lichtenstein,	Phys. Rev. B {\bf 105}, 155151 (2022).
\bibitem{OurMott} T. B. Mazitov, A. A. Katanin, Phys. Rev. B 105, L081111 (2022).


\bibitem{OurMott1} E. G. C. P. van Loon, F. Krien, and A. A. Katanin,
Phys. Rev. Lett. {\bf 125}, 136402 (2020).
\bibitem{Triangular1} A. Vrani\'{c}, J. Vu\v{c}i\v{c}evi\'{c}, J. Kokalj, J. Skolimowski, R. \v{Z}itko, J. Mravlje, and D. Tanaskovi\'{c},
Phys. Rev. B {\bf 102}, 115142 (2020).


%
\bibitem{NRG} R. \v{Z}itko and T. Pruschke,
Phys. Rev. B {\bf 79}, 085106 (2009).
\bibitem{TRIQS} O. Parcollet, M. Ferrero, T. Ayral, H. Hafermann, I. Krivenko, L. Messio, P. Seth,  Comp. Phys. Comm. {\bf 196}, 398 (2015); https://triqs.github.io/nrgljubljana\_interface/.

\bibitem{Pruschke} M. Jarrell and Th. Pruschke,  Z. Phys. B {\bf 90}, 187 (1993).

\bibitem{Log} T.-F. Fang, N.-H. Tong, Z. Cao, Q.-F. Sun, and H.-G. Luo, Phys. Rev. B {\bf 92}, 155129 (2015).

\bibitem{iQIST} L. Huang, Y. Wang, Z. Y. Meng, L. Du, P. Werner, and X. Dai, 
Comput. Phys. Commun. \textbf{195}, 140 (2015); L. Huang, \textit{ibid.} \textbf{221}, 423 (2017).

\bibitem{OurVanadium} A. S. Belozerov, A. A. Katanin, V. I. Anisimov, arXiv:2206.12383.

\bibitem{Wilson} K. Wilson, Rev. Mod. Phys. {\bf 47}, 773 (1975).

\bibitem{Wilson1} H. R. Krishna-murthy, J. Wilkins, and K. G. Wilson, Phys. Rev. B {\bf 21}, 1003 (1980).

\bibitem{Melnikov} V. I. Mel'nikov, Soviet Phys. JETP Lett. {\bf 35}, 511 (1982).

\bibitem{Tsvelik} A. M. Tsvelick and P. B. Wiegmann, Adv. Phys. {\bf 32}, 453 (1983).

\bibitem{OurFe1}  A. A. Katanin,
A. I. Poteryaev, A. V. Efremov, A. O. Shorikov, S. L. Skornyakov, M. A. Korotin, and V. I. Anisimov, 
Phys. Rev. B {\bf 81}, 045117 (2010).

\bibitem{OurFeGamma} P. A. Igoshev, 
A. V. Efremov, A. I. Poteryaev, A. A. Katanin, and V. I. Anisimov, Phys. Rev. B {\bf 88}, 155120 (2013).

\bibitem{Horvatic} B. Horvati\'{c}, D. \v{S}ok\v{c}evi\'{c}, V. Zlati\'{c}, Phys. Rev. B {\bf 36}, 675 (1987).

\bibitem{Hewson} A. C. Hewson, The Kondo Problem to Heavy Fermions, Cambridge University Press, 1993.

\bibitem{Varma} C. M. Varma, P. B. Littlewood, S. Schmitt-Rink, E. Abrahams, and A. E. Ruckenstein
Phys. Rev. Lett. {\bf 63}, 1996 (1989).

\bibitem{Lavagna} 
M. Lavagna, C. Lacroix, and M. Cyrot, J. Phys. F {\bf 12}, 745 (1982).

\bibitem{CompAnd} G-M Zhang and A. C. Hewson, Phys. Rev. Lett. {\bf 76}, 2137 (1996).

\bibitem{Castellani} C. Castellani, C. Di Castro, D. Feinberg, and J. Ranninger
Phys. Rev. Lett. {\bf 43}, 1957 (1979).

\bibitem{Hirsch} J. E. Hirsch,
Phys. Rev. Lett. {\bf 54}, 1317 (1985).

\bibitem{Cyrot} M. Cyrot, Sol. St. Comm. {\bf 60}, 253 (1986).

\bibitem{OurRev} G. Rohringer, H. Hafermann, A. Toschi, A. A. Katanin, A. E. Antipov, M. I. Katsnelson, A. I. Lichtenstein, A. N. Rubtsov, and K. Held, Rev. Mod. Phys. {\bf 90}, 025003 (2018).

\bibitem{Yang} Z. Long, J. Wang and Yi-feng Yang, {Phys. Rev. B {\bf 106}, 195128 (2022).}



\end{thebibliography}
\end{document}